\documentclass[]{spie}  

\usepackage{amsmath,amsfonts,amssymb}
\usepackage{graphicx}
\usepackage{epstopdf}
\usepackage[colorlinks=true, allcolors=blue]{hyperref}

\newcommand{\Der}{\mathrm{d}}
\newcommand{\Img}{\mathrm{i}}
\newcommand{\Eul}{\mathrm{e}}

\title{Approaching the ultimate capacity limit in deep-space optical communication}

\author[a,b]{Konrad Banaszek}
\author[a,b]{Ludwig Kunz}
\author[a]{Marcin Jarzyna}
\author[a,b]{Micha{\l} Jachura}

\affil[a]{Centre for Quantum Optical Technologies, University of Warsaw, Banacha 2c, 02-097~Warszawa, Poland}
\affil[b]{Faculty of Physics, University of Warsaw, Pasteura 5, 02-093 Warszawa, Poland}

\authorinfo{Further author information: (Send correspondence to K.B.)\\K.B.: E-mail: k.banaszek@cent.uw.edu.pl}

\begin{document}
\maketitle

\begin{abstract}
The information capacity of an optical channel under power constraints is ultimately limited by the quantum nature of transmitted signals. We discuss currently available and emerging photonic technologies whose combination can be shown theoretically to enable nearly quantum-limited operation of a noisy optical communication link in the photon-starved regime, with the information rate scaling linearly in the detected signal power. The key ingredients are quantum pulse gating to facilitate mode selectivity, photon-number-resolved direct detection, and a photon-efficient high-order modulation format such as pulse position modulation, frequency shift keying, or binary phase shift keyed Hadamard words decoded optically using structured receivers.
\end{abstract}


\section{INTRODUCTION}
\label{Sec:Introduction}

Quantum mechanics defines the ultimate capacity limits of optical communication links.\cite{GiovannettiGarciaPatronNPH2014} In recent years, a good deal of effort has been put into devising schemes that would attain this limit in the photon-starved regime, when the propagating optical signals experience substantial attenuation bringing the number of detected photons per slot to much less than one on average.\cite{ErkmenMoisionSPIE2010,GuhaPRL2011,ErkmenMoisionSPIE2012,TakeokaGuhaPRA2014,RosatiMariPRA2016,ChungGuhaPRA2017} Such a regime is routinely encountered in implemented and planned optical links for transferring data collected by scientific instruments onboard deep-space missions.\cite{HemmatiBiswasProcIEEE2011,MoisionIPNPR2014,BorosonSPIE2018,ArapoglouSPIE2018}

In this contribution, we address the role of excess optical noise, which adds a random fluctuating component to the complex amplitude of the transmitted signal. Such noise originates e.g.\ from stray light scattering in a free-space optical channel. We point out that the excess noise defines the ultimate quantum mechanical limit on the photon information efficiency attainable in deep-space optical communication systems. Interestingly, this bound has a very weak, logarithmic dependence on the excess noise power spectral density. We discuss assumptions underlying its derivation and identify emerging photonic technologies that could make it relevant to the operation of actual deep-space optical communication links.

\section{CAPACITY LIMITS}
\label{Sec:limits}
Throughout this paper we will assume that the time axis is divided into discrete slots. The duration $\tau$ of a single slot is defined by the inverse of the bandwidth $B$ of the optical channel, $\tau=1/B$. A convenient figure of merit is the average optical energy detected in a single slot, expressed in the units of the energy $hf_c$ of a single photon at the carrier frequency $f_c$. Here $h= 6.626 \times 10^{-34}~{\textrm J} \cdot {\textrm s}$ is Planck's constant. The resulting quantity is the {\em average detected photon number per slot}
\begin{equation}
n_a = \frac{\eta P_{\textrm{tx}} \tau}{h f_c} = \frac{\eta P_{\textrm{tx}}}{ B h f_c}
\label{Eq:nadef}
\end{equation}
where $P_{\textrm{tx}}$ is the trasmitter power and $\eta$ is a linear factor characterizing the overall channel transmission. For notational simplicity we will take $\eta$ to incorporate also the efficiency of the detection system. If in the course of propagation the signal acquires a stochastic component generated by broadband Gaussian noise with power spectral density $N_b$, its contribution can be quantified using the {\em average number of background photons detected per slot},
\begin{equation}
n_b = \frac{N_b}{hf_c}.
\label{Eq:nbdef}
\end{equation}

The information capacity of an optical channel is typically analyzed under the assumption of a fixed average power.
If both field quadratures are utilized to encode information and are read out via shot-noise limited heterodyne detection, the capacity per slot takes the form
\begin{equation}
\label{Eq:Chet}
{\sf C}_{\text{het}} = \log_2 \left( 1 + \frac{n_a}{1+n_b}\right)
\end{equation}
and the optimal constellation in the complex amplitude plane is given by a phase-invariant two-dimensional Gaussian distribution centered at zero. When information is encoded only in one field quadrature and retrieved using a homodyne receiver operating at the shot-noise limit, the capacity per slot reads
\begin{equation}
\label{Eq:Chom}
{\sf C}_{\text{hom}} = \frac{1}{2} \log_2
\left( 1 + \frac{4n_a}{1+2n_b}\right).
\end{equation}
In this case the optimal constellation is one-dimensional with the Gaussian distribution for the detected quadrature
and the conjugate quadrature set to zero.

Expressions given in Eq.~(\ref{Eq:Chet}) and Eq.~(\ref{Eq:Chom}) follow from the Shannon-Hartley theorem. The fractions appearing in the argument of the logarithm have the straightforward interpretation of the signal-to-noise ratio. In each case the noise term in the denominator is a sum of two contributions. The first one, equal to one owing to the choice of units made here, comes from the detection process assumed to operate at the shot noise limit.\cite{YuenChanOptLett1983,YuenChanOptLett1983Err} The second contribution stems from the excess noise added to the signal propagating through the channel. The shot noise  defines natural units for the signal and the background noise strength used in definitions (\ref{Eq:nadef}) and (\ref{Eq:nbdef}).

The ultimate quantum mechanical capacity\cite{Holevo1973} is derived under the assumption that at the output of the channel one can implement the most general measurement permitted by quantum theory. We will refer to this quantity as the {\em Holevo capacity}. The explicit expression for the Holevo capacity in the case of a noisy channel reads:\cite{GiovannettiGarciaPatronNPH2014}
\begin{equation}
\label{Eq:CHol}
{\sf C}_{\text{Hol}} = g(n_a + n_b) - g(n_b), \qquad g(x) = (x+1) \log_2 (x+1) - x \log_2 x.
\end{equation}
The above expression has a different mathematical form compared to Eqs.~(\ref{Eq:Chet}) and (\ref{Eq:Chom}). Somewhat bafflingly from the perspective of the classical theory, the signal-to-noise ratio term can no longer be identified in a straightforward manner.

\begin{figure}
\centering
\includegraphics[width=0.8\linewidth]{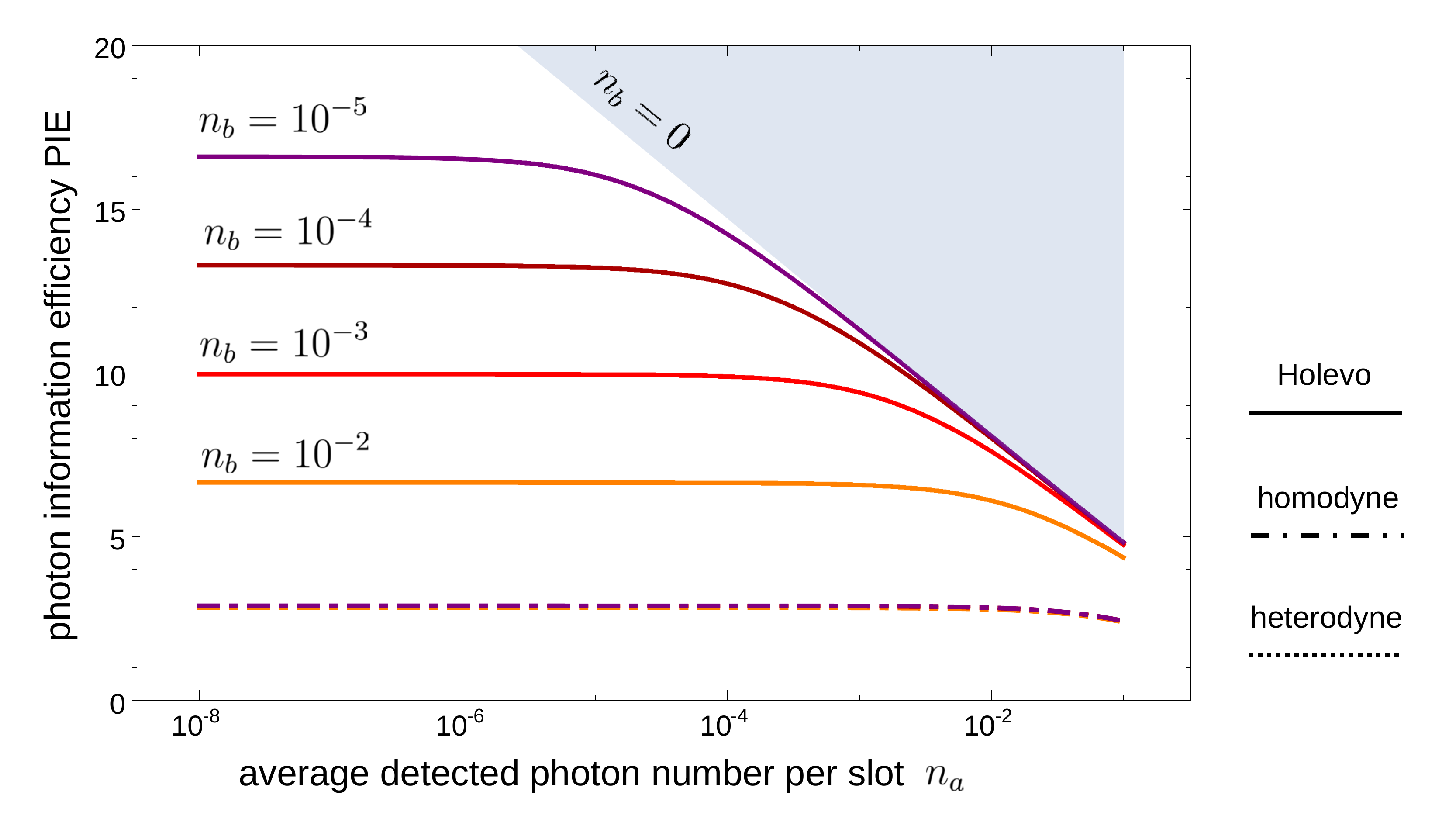}
\caption{Capacity limits in photon-starved communication shown as a dependence of the photon information efficiency PIE on the average number of detected photons per slot for several values of the background noise strength $n_b=10^{-5}, 10^{-4}, 10^{-3}$, and $10^{-2}$. Solid lines depict the Holevo bound, while overlapping sets dotted and dash-dotted lines correspond respectively to heterodyne and homodyne capacities. The edge od the shaded area indicates the noiseless Holevo bound given by $g(n_a)$ as specified in Eq.~(\ref{Eq:CHol}).}\label{Fig:PIENoisy}
\end{figure}

In the following,  it will be convenient to switch to photon information efficiency (PIE) defined by ${\sf PIE} = {\sf C}/n_a$ as a performance measure. PIE provides the proportionality factor between the attainable information rate and the detected photon flux. Fig.~\ref{Fig:PIENoisy} depicts PIE in the photon-starved regime for several values of the background noise in the range $10^{-5} \le n_b \le 10^{-2}$. A dramatic difference is seen in the asymptotic limit $n_a \rightarrow 0$ between the conventional coherent (heterodyne or homodyne) detection on one hand and the Holevo bound on the other hand. Analytical expressions for the asymptotic values of PIE can be easily obtained by expanding respective capacities up to the linear term in $n_a$, which yields:
\begin{equation}
{\sf PIE}_{\text{het}} \rightarrow \frac{1}{1 + n_b}\log_2\Eul, \qquad
{\sf PIE}_{\text{hom}} \rightarrow \frac{2 }{1 + 2 n_b}\log_2\Eul, \qquad
{\sf PIE}_{\text{Hol}} \rightarrow \log_2 \left( 1 + \frac{1}{n_b} \right).
\end{equation}
For heterodyne and homodyne detection, PIE is upper bounded respectively by $1~\text{nat} = \log_2\Eul$ and $2~\text{nats}$ of information. Furthermore, as long as $n_b \ll 1$ the background noise plays a negligible role, being dominated by the detection shot noise which remains at a constant level. In contrast, the Holevo bound for PIE depends dramatically on the background noise strength. In the regime $n_b \ll 1$ it exhibits a simple logarithmic dependence ${\sf PIE}_{\text{Hol}} \sim \log_2 (1/n_b)$. For example, $n_b = 10^{-3}$ sets the effective upper bound on PIE slightly at less than 10 bits per photon. Based on Fig.~\ref{Fig:PIENoisy} one can infer a simple rule of thumb that the asymptotic value of the Holevo PIE bound is reached when the signal strength $n_a$ becomes lower than the background noise strength, i.e.\ $n_a \lesssim n_b$.

\section{Noise rejection}
\label{Sec:NoiseRejection}

\begin{figure}
\centering
\includegraphics[width=0.95\linewidth]{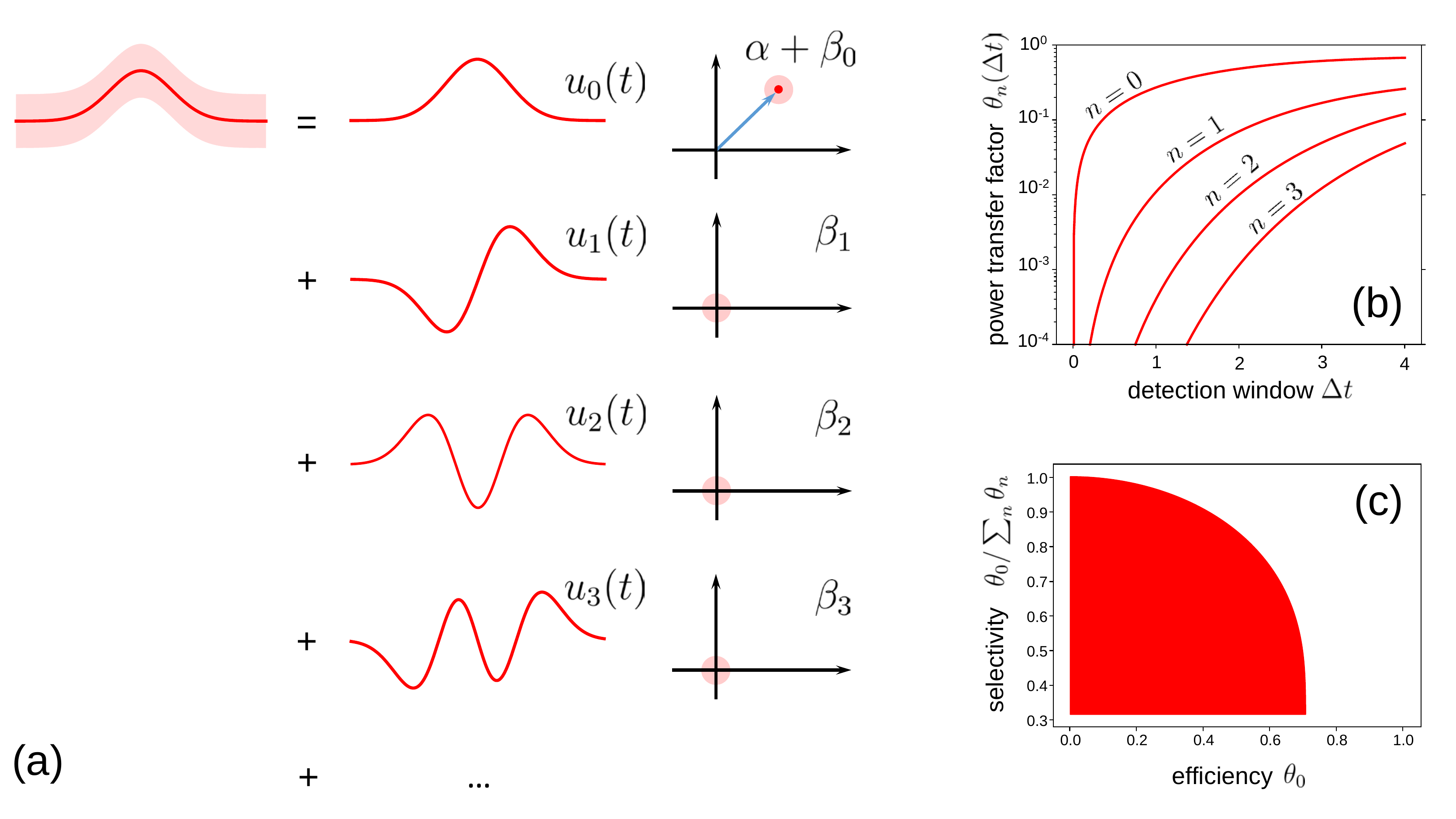}

\vspace{5mm}

\caption{(a) Noisy signal field can be decomposed in a basis of temporal modes. The waveform $u_0(t)$ carrying information in the complex amplitude $\alpha$ is complemented with orthogonal modes $u_1(t), u_2(t), u_3(t), \ldots$. The noise contributes random complex amplitudes $\beta_0, \beta_1, \ldots$ to individual modes.
(b) The power transfer coefficients $\theta_n(\Delta t)$ for Gauss-Hermite modes $u_n(t)$ as a function of the detection window $\Delta t$ with the matched filter implemented optically in the form of spectral optical filtering with peak transmission equal to one. (c) The resulting trade-off between the filter efficiency $\theta_0$ and the selectivity $\theta_0/\sum_{n=0}^{\infty} \theta_n$ with the accessible values shown as the solid red area.
\label{Fig:Noise}}
\end{figure}

It is illuminating to discuss in more detail the noise model underlying the capacity limits presented in the preceding section. The optical field emerging from a noisy channel, shown schematically in Fig.~\ref{Fig:Noise}(a), carries information in the complex amplitude $\alpha$ of a certain normalized signal mode (waveform) characterized by its envelope $u_0(t)$. Consider broadband additive Gaussian excess noise, visualized in Fig.~\ref{Fig:Noise}(a) as a semi-transparent strip overlaid on the signal waveform. The noise can be decomposed in an basis orthonormal of modes $u_0(t), u_1(t), u_2(t), \ldots$ shown in Fig.~\ref{Fig:Noise}(a), chosen such that the first mode $u_0(t)$ is identical with the signal mode. Thus the signal mode carries the modulated amplitude plus the noise contribution, while all the other modes contain only the background noise. A single realization of the field can be written as ${\cal E}(t) \exp(-2 \pi \Img f_c t)$ with the envelope given by a sum
\begin{equation}
{\cal E}(t) \propto (\alpha+\beta_0) u_0(t) + \beta_1 u_1(t) + \beta_2 u_2(t) + \ldots .
\label{Eq:E(t)=sum}
\end{equation}
We will choose the units for the complex amplitudes multiplying mode functions such that their squared absolute values specify the average numbers of photons in individual modes.
For broadband noise, the stochastic amplitudes $\beta_0, \beta_1, \beta_2, \ldots$ are characterized by the same phase-invariant Gaussian distribution with a zero mean, the second moment equal to
\begin{equation}
\langle |\beta_0|^2 \rangle = \langle |\beta_1|^2 \rangle = \langle |\beta_2|^2 \rangle = \ldots = n_b,
\end{equation}
and lack of correlations, $\langle \beta_i \beta_j \rangle = \langle \beta_i \beta_j^\ast \rangle = 0$ for $i\neq j$.

Expressions for channel capacities presented in Eqs.~(\ref{Eq:Chet})--(\ref{Eq:CHol}) are derived assuming that $\langle |\alpha|^2 \rangle = n_a$ and that only the noise present in the signal mode $u_0(t)$ is taken into account, while contributions from modes $u_1(t), u_2(2), \ldots$ are rejected. This assumption is naturally satisfied for coherent detection, which measures only the amplitude of a specific waveform matching the temporal shape of the local oscillator.\cite{YuenShapiroIEEE-IT1980} Alternatively, the detected mode can be defined in digital signal postprocessing if the detector electronic output is sampled at a sufficiently high rate. On the other hand, the applicability of such a model to the direct detection scenario requires a more careful inspection.

According to the photodetection theory,\cite{MandelWolfSemiclPhot}
the photocount statistics is a function of the optical power $|{\cal E}(t)|^2$ time-integrated over the detection window. Clearly, with ${\cal E}(t)$ given by Eq.~(\ref{Eq:E(t)=sum}) such an integral will contain contributions not only from the mode $u_0(t)$, but also from other modes $u_1(t), u_2(t), \ldots$. In principle, the contribution to the optical field from the signal mode $u_0(t)$  in Eq.~(\ref{Eq:E(t)=sum}) can be separated using the {\em matched filter}\cite{LapidothBook2017} $f(t) \propto u_0^\ast(-t)$ such that the convolution
\begin{equation}
(f \star {\cal E})(t) =
 \int_{-\infty}^{\infty} \Der t' \, f(t - t') {\cal E}(t')
\end{equation}
yields $(f \star {\cal E})(0) \propto \alpha+ \beta_0$ at time $t = 0$. Optically, convolution in the temporal domain can be realized by sending the signal through a spectral filter with the amplitude transfer function given by the Fourier transform of $f(t)$. Detecting subsequently the output field over a short time window centered at $t=0$ would provide response dependent only on the amplitude $\alpha + \beta_0$ of the signal mode. However, the basic drawback of this scheme is the loss of the signal power. This can be seen most easily by inspecting the effective power transfer coefficients $\theta_n(\Delta t)$ that specify how much optical power from individual modes $u_n(t)$ contributes to the detection window of duration $\Delta t$:
\begin{equation}
\theta_n(\Delta t) = \int_{-\Delta t/2}^{\Delta t/2} \Der t \, |(f \star u_n)(t)|^2.
\end{equation}
Fig.~\ref{Fig:Noise}(b) shows the coefficients $\theta_n(\Delta t)$ assuming Gauss-Hermite modes $u_n(t) = H_n(t) e^{-t^2/2} / (\pi^{1/2} 2^n n!)^{1/2}$ and unit peak transmission of the spectral filter. It is seen that while the selectivity improves with shorter time windows $\Delta t$, the efficiency of selecting the signal component drops significantly below $100\%$. Importantly, even for long time windows $\Delta t$ the power transfer is limited to slightly below $71\%$. The resulting trade-off between the efficiency $\theta_0$ and the selectivity $\theta_0/\sum_{n=0}^{\infty} \theta_n$ is depicted in Fig.~\ref{Fig:Noise}(c).

The capacity limits defined in Eqs.~(\ref{Eq:Chet})--(\ref{Eq:CHol}) involve an absolute, input-independent scale for the signal and noise strengths. This scale is defined by the detector shot-noise level in the case of coherent detection, or more fundamentally by the granular nature of electromagnetic radiation taken into account in the Holevo bound. Consequently, to approach the capacity limit it would be necessary to pick up the signal mode with 100\% efficiency {\em and}\/ 100\% selectivity.

Deficiencies of conventional spectral filtering described above are in principle absent in the recently introduced technique of quantum pulse gating \cite{EcksteinBrechtOpEx2011,BrechtReddyPRX2015}. The basic idea is to select from a composite field, such as the one given in Eq.~(\ref{Eq:E(t)=sum}), a single temporal mode with $100\%$ efficiency in a way that does not alter its quantum statistical properties. Such a quantum pulse gate (QPG) can be implemented using three-wave mixing in a $\chi^{(2)}$ nonlinear medium with carefully engineered phase matching properties. The composite field is sent into the medium along with an auxiliary pulse so that sum-frequency generation takes place only between the mode of interest and the auxiliary pulse mode. At the output of the medium the frequency-upconverted mode is separated spectrally, using e.g.\ a dichroic mirror, from the remaining field. In a proof-of-principle demonstration with near-infrared ultrashort sub-picosecond pulses,\cite{ReddyRaymerOpEx2017} conversion efficiency exceeding $80\%$ for the Gauss-Hermite signal mode $u_0(t)$ has been achieved with less than $20\%$ power transferred for the first-order mismatched mode $u_1(t)$. While theoretical limits on the conversion efficiency in a single-stage QPG have been indicated,
the coherent nature of the sum-frequency generation process allows one to overcome them by by cascading two or more QPGs with individual conversion efficiencies lower than one.\cite{ReddyRaymerOptLett2014,ReddyRaymerOptica2018} It is worth to mention also other functionality of quantum pulse gating, such as changing the bandwidth of the gated signal\cite{AllgaierAnsariNatComm2017} and converting the carrier frequency to the spectral range where more efficient photon counting is available.

\section{Pulse position modulation}

The standard strategy to achieve high PIE in photon-starved optical communication is to employ the pulse position modulation (PPM) format with direct detection. Photon counting as a detection technique goes beyond the measurement of field quadratures, which makes void capacity limits resulting from the application of the Shannon-Hartley theorem to readout based on conventional coherent detection.

In a typical ground receiver setup for dowlink transmission the photon counting detector responds to radiation present in multiple temporal and spatial modes. In the temporal degree of freedom, the effective number of modes is roughly given the product of the slot duration and the bandwidth of the spectral filter used to suppress the background noise outside the signal spectrum.\cite{LandauPollakBellSys1962} In the spatial degree of freedom, atmospheric turbulence distorts the signal wavefront well beyond the diffraction limit. The distorted wavefront needs to be accommodated with a sufficiently large active area of the detector that consequently becomes sensitive to background radiation present in spatial modes other than the signal one. When background counts are generated by noise contributions from a large number of weakly excited modes, their statistics can be modelled by a Poissonian distribution. The PPM limits for such a multimode noise model, assuming a photon counting detector operated in the Geiger mode, i.e.\ providing a binary response whether at least one photon has been detected in a given slot or none at all, have been recently analyzed by Zwoli\'{n}ski {\em at al.}\cite{ZwolinskiJarzynaOpEx2018} The basic conclusion is that for a fixed background noise level one can in principle achieve non-zero photon information efficiency in the limit of a diminishing signal strength. The prerequisites are complete decoding to recover information from all sequences of photocounts that may occur within individual PPM frames and sufficient concentration of optical energy in signal pulses, which should contain between $\sim 0.2$ and $\sim 1.1$ photon at the detection stage for noise strengths ranging from $10^{-5}$ to $10^{-1}$ background counts per slot.\cite{JarzynaICSO2018}

Obviously, the multimode noise model cannot be directly compared with the Holevo capacity bound. From the theoretical viewpoint, background noise present in modes other than the signal mode can be rejected without affecting the signal itself and therefore it does not fundamentally limit the link capacity. Although achieving in practice single-mode direct detection of the signal waveform would be challenging,  it may no longer seem totally outlandish on the second thought. Adaptive optics technology enables compensation of wavefront distortion effects and could facilitate delivery of the received signal to the detection stage in a single spatial mode. Further, as discussed in Sec.~\ref{Sec:NoiseRejection}, the technique of quantum pulse gating provides in principle means to pick up the temporal signal mode with nearly 100\% efficiency and very good selectivity.

Given the above motivation, it is interesting to examine the efficiency of a PPM link assuming the single-mode model for the background noise. In this scenario, the probabilities of registering $k$ photocounts on a photon number resolving detector respectively for an empty slot ``0'' and a pulse ``1'' are given by:\cite{ArecchiBerneIEEE-QE1966}
\begin{equation}
p_0 (k) = \frac{n_b^k}{(1+n_b)^{k+1}} , \qquad p_1(k) = \frac{n_b^k}{(1+n_b)^{k+1}} \exp\left( - \frac{M n_a}{1+n_b}\right)
L_k \left( - \frac{M n_a}{n_b (1+n_b)}\right).
\end{equation}
Here $M$ is the PPM order and $L_k(\cdot)$ denotes the $k$th Laguerre polynomial. For a given average detected signal photon number $n_a$ and the background noise strength $n_b$ optimization has been carried out over the PPM order $M$ taking as the cost function a lower bound on the PPM efficiency based on relative entropy. Two types of direct detection have been considered: photon number resolved (PNR) detection, which gives the actual photocount number $k$, and Geiger-mode detection, which discriminates only between $k=0$ and $k \ge 1$.

\begin{figure}
\centering
\includegraphics[width=\linewidth]{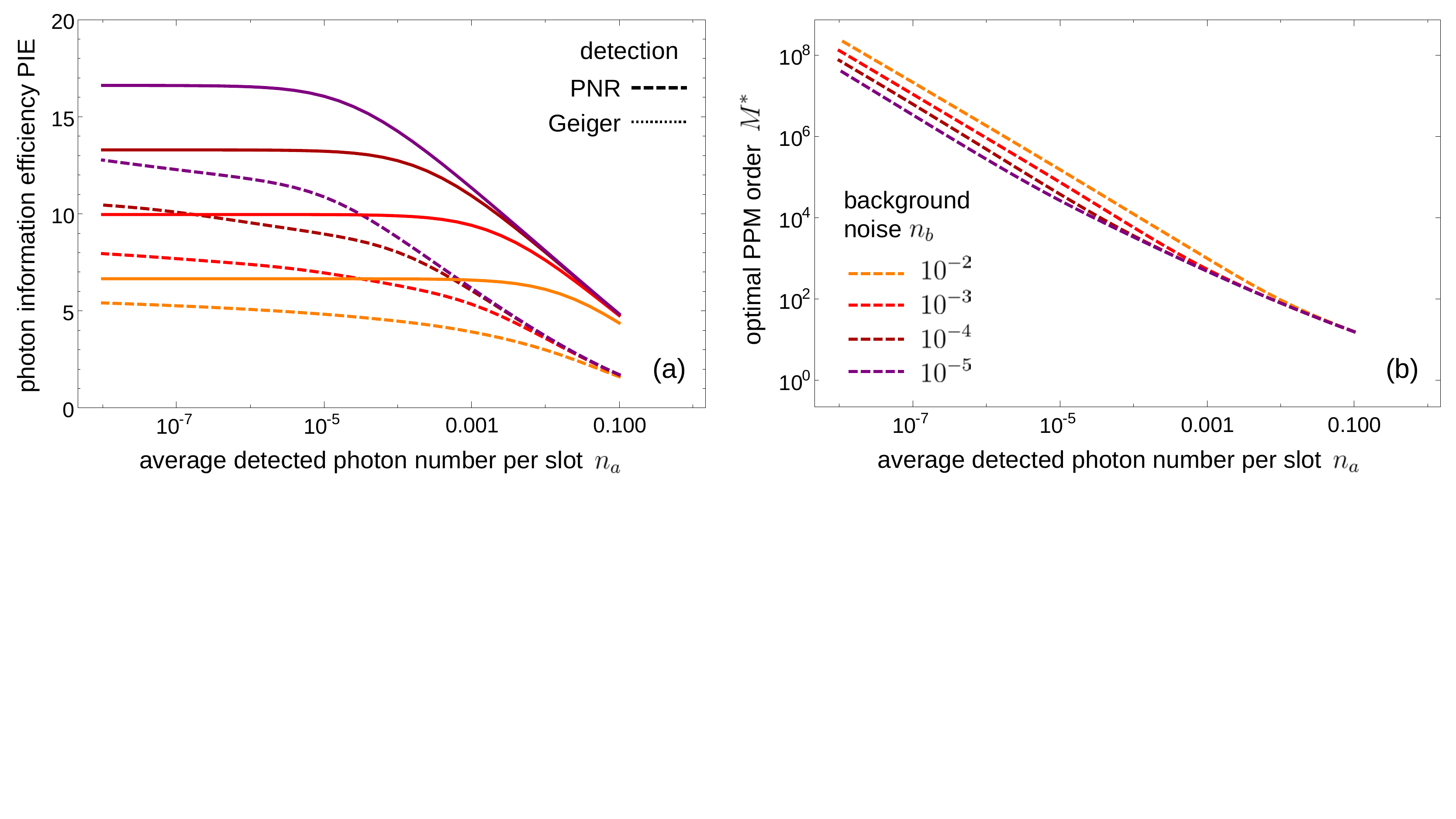}

\vspace*{-3.5cm}

\caption{(a) Photon information efficiency for the PPM format as a function of the average detected photon number $n_a$ optimized in the case of photon number resolved (dashed lines) and Geiger-mode (thin dotted lines) detection. For reference, solid lines depict the Holevo bound. (b) Optimal PPM orders for photon number resolved detection as a function of the signal strength.\label{Fig:PPM}}
\end{figure}

Fig.~\ref{Fig:PPM}(a) depicts results of optimization for background noise strengths $n_b =10^{-5}, 10^{-4}, 10^{-3}$, and $10^{-2}$. The most interesting feature is that for PNR detection the gap between the optimized PPM photon information efficiency and the Holevo bound closes with diminishing signal strength. Indeed, it can be shown mathematically that the capacity per unit cost for PPM with PNR detection approaches the Holevo limit.\cite{JarzynaUnpublished} Photon number resolving capability is essential here: it is seen that for Geiger-mode detection the attainable PIE stays below the Holevo bound. This result is in a stark contrast with the noiseless ($n_b=0$) scenario, where a double-logarithmic gap opens up with the diminishing signal strength between optimized PPM PIE and the Holevo bound.\cite{KochmanWangTIT2014,JarzynaKuszajOPEX2015,JarzynaBanaszekICSOS2017}

\section{Outlook}
\label{Sec:Outlook}

Achieving information efficiency identified in the preceding section requires implementation of the PPM format with very high orders, as shown in Fig.~\ref{Fig:PPM}(b). Basically, the pulses should be bright enough to generate several photocounts on the detector to facilitate robust discrimination against background counts. This detection regime resembles that employed recently in low-intensity LiDAR technology,\cite{EisenbergSPIE2018} where the light intensity in the few-photon range also needs to be measured within a short time window. Another prerequisite to approach the Holevo limit is soft-decoding of the receiver output with efficiency attaining the Shannon information limit, which is implicitly assumed in the relative-entropy bound used in numerical calculations.

\begin{figure}
\centering
\includegraphics[width=\linewidth]{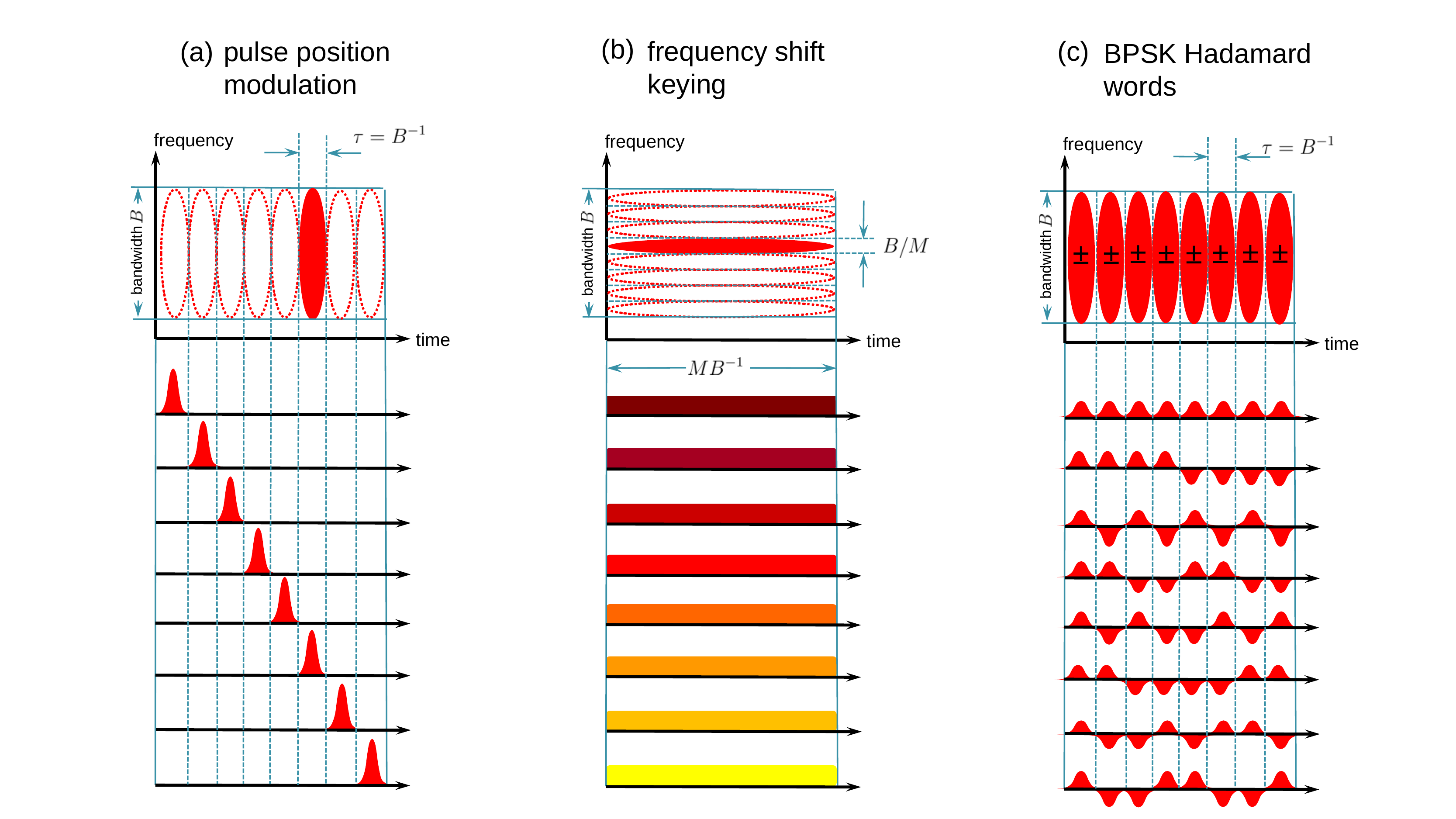}

\bigskip
\caption{Time-frequency diagrams for (a) pulse position modulation, (b) frequency shift keying, and (c) Hadamard words composed from the BPSK alphabet shown for the format order $M=8$. In all three cases individual symbols occupy the same $MB^{-1} \times B$ area in the time-frequency plane and are characterized by the optical energy $Mn_a$.\label{Fig:TimeFrequencyDiagram}}
\end{figure}

The principal technological challenge on the transmitter side to produce a high-order PPM signal is the high peak-to-average power ratio of the laser source. This obstacle could be removed by switching to one of other high-order modulation formats visualized using time-frequency diagrams presented in Fig.~\ref{Fig:TimeFrequencyDiagram} that are capable of delivering equivalent information efficiency.\cite{BanaszekICSO2018} In the case of the PPM format depicted in Fig.~\ref{Fig:TimeFrequencyDiagram}(a), the modulation bandwidth $B$ defines the length $\tau$ of a single slot in the time domain as $\tau = B^{-1}$. Therefore one $M$-ary PPM symbol has duration $M\tau = MB^{-1}$, covers the area $MB^{-1}$ (time) $\times$ $B$ (frequency), and carries optical energy $Mn_a$ measured at the detection stage. As illustrated in Fig.~\ref{Fig:TimeFrequencyDiagram}(b), the same time-frequency area can be sliced in the spectral domain into $M$ frequency slots, each of spectral width $B/M$ and temporal duration $MB^{-1}$, which yields the frequency shift keying (FSK) modulation format.
The optical energy $Mn_a$ of each symbol is now spread evenly across the entire time frame.
In the context of free-space optical communication, scalable FSK modulation has been demonstrated by selecting individual spectral lines from a frequency comb using a sequence of electrooptically controlled Mach-Zehnder interferometric filters.\cite{SavageRobinsonOpEx2013} On the other hand, the readout of FSK symbols would require a high-resolution, low-loss spectrometer followed by an array of photon couting detectors.

Yet another option to prepare $M$ symbols covering the $MB^{-1}\times B$ time-frequency area with a uniform distribution of instantaneous power is to generate sequences of binary phase shift keyed (BPSK) pulses in $M$ consecutive time slots, as shown in Fig.~\ref{Fig:TimeFrequencyDiagram}(c). When signs for BPSK sequences are taken as rows of Hadamard matrices, it is possible to convert all-optically received pulse trains into the PPM format using structured receivers that implement optical interference between multiple time slots.\cite{GuhaPRL2011} Importantly, scalable designs for structured optical receivers have been presented.\cite{BanaszekJachuraICSOS2017} A major advantage of the modulation format based on such BPSK Hadamard words is that pulse sequences could be generated by a transmitter based on standard telecom components. Furthermore, the modulation order $M$ is software-defined in the encoding layer of the communication system and changing it does not require any hardware adjustments. The main challenge of implementing this modulation format would be the construction of a reliable structured optical receiver. Elements of the required technology are shared with other setups for free-space optical communication, such as links utilizing the differential phase shift keying (DPSK) format\cite{SodnikSansICSOS2012} and quantum key distribution based on time-bin qubit encoding.\cite{JinAgnePRA2018} Looking at less mature technologies, the BPSK signal might be a good candidate to implement noise rejection using the quantum pulse gating technique, provided that temporal synchronization with the receiver setup can be established.

\acknowledgments

We acknowledge insightful discussions with C. Antonelli, H. Eisenberg, A. Mecozzi, and M. Shtaif.
This work is part of the project ``Quantum Optical Communication Systems'' carried out within the TEAM
programme of the Foundation for Polish Science co-financed by the European Union under the European
Regional Development Fund.

\bibliography{deepspace}

\bibliographystyle{spiebib}

\end{document}